# Influence of $CeO_2MnOx$ heterostructure on Hydrogen Peroxide Electrogeneration on Carbon-Based Catalysts

Caroline de O. Carrilho[1], Juliana M. S. de Jesus[1], João Paulo C. Moura[1], Dara Silva Santos[1], Aline B. Trench[1], Caio Machado Fernandes[1], Aila O. Santos[2], Odivaldo C. Alves[2], Júlio C. M. Silva[2], Mauro C. dos Santos[1*]

*[1]Laboratory of Electrochemistry and Nanostructured Materials (LEMN) - Center for Natural and Human Sciences (CCNH), Federal University of ABC (UFABC), CEP 09210-170, Avenida dos Estados, Bairro Bangu, Santo André, São Paulo, Brazil*

*[2]Department of Physical Chemistry, Fluminense Federal University, Campus Valonguinho, 24020-141 Niterói, Rio de Janeiro, Brazil*

Corresponding Author:

*E-mail: mauro.santos@ufabc.edu.br (M.C. Santos).

## ABSTRACT

The sustainable electrogeneration of hydrogen peroxide ($H_2O_2$) via the two-electron oxygen reduction reaction ($2e^-$ ORR) represents a promising alternative to conventional production methods. In this study, $CeO_2$ and $CeO_2MnOx$ nanoparticles were synthesized and supported on Vulcan XC-72 carbon at varying loadings (1, 3, and 5%), aiming to assess the lowest metal loading and high $H_2O_2$ electrosynthesis. Physicochemical characterizations confirmed the successful formation of $CeO_2$ nanowires and the effectiveness of the MnOx surface modification. XRD, TEM, XPS, EPR, and contact angle analyses revealed that $CeO_2$ loading increased surface hydrophilicity by the presence of oxygenated functional groups, thereby favoring electrochemical activity. On the other hand, all $CeO_2MnO_x$ loadings were statistically equal to Vulcan XC-72 in terms of contact angle. Electrochemical evaluations using a rotating ring-disk electrode (RRDE) demonstrated enhanced ORR activity and high $H_2O_2$ selectivity for the 1% $CeO_2MnOx/C$ and 3% $CeO_2/C$ catalysts, achieving up to 90% selectivity and elevated ring currents. The results suggest that low metal loading and surface modification via MnOx improve the balance between active site exposure, oxygen adsorption, and intermediate stabilization, thus favoring the selective $2e^-$ pathway. These findings support the development of cost-effective, non-noble-metal catalysts for green $H_2O_2$ via electrosynthesis.

## 1. INTRODUCTION

Electrogeneration of hydrogen peroxide ($H_2O_2$) provides a sustainable, decentralized green oxidant [1–5]. Unlike the energy-intensive anthraquinone process, electrochemical synthesis allows on-site, on-demand production using water, oxygen, and renewable electricity, reducing emissions and transport needs [1–5]. $H_2O_2$ decomposes into water and oxygen, leaving no pollutants, and is used in advanced oxidation for water treatment, degrading contaminants and pathogens. Electrogenerated $H_2O_2$ offers an eco-friendly, cost-effective alternative for sustainable water management and pollution control [1–5]. The oxygen reduction reaction (ORR) powers energy devices such as fuel cells and batteries and enables peroxide synthesis by converting oxygen into water ($H_2O$) or $H_2O_2$. Its slow kinetics require active, stable catalysts to enhance efficiency and selectivity, reduce overpotentials, and enable sustainable energy and green chemistry under mild conditions [1–5].

Recent reviews highlight advances in understanding the ORR mechanism and developing new electrocatalysts using synthesis strategies, *in-situ* characterization, and computational tools, offering an overview of catalyst design, including the role of adsorption energies for intermediates (*OOH, *O, *OH), machine learning for ORR activity prediction, and in situ electrochemical techniques to study active-site dynamics [6–8]. These advances have improved noble metal, transition-metal, single-atom, and carbon-based catalysts for water- and $H_2O_2$-producing ORR pathways [6,7,9].

Designing selective, efficient electrocatalysts for two-electron reactions involves tuning key properties. The catalyst should have intermediate binding strength for reaction intermediates, such as *OOH, in oxygen reduction to $H_2O_2$, stabilizing the intermediate to promote its formation without further transfer, thus favoring the two-electron pathway [6,7,10]. Balancing adsorption energies directs the reaction toward the desired product.

Additionally, adjusting the electronic structure of the active sites, such as the d-band center and surface electronic density, reduces the kinetic barriers for the two-electron process and suppresses the four-electron pathway. Surface properties such as hydrophobicity or hydrophilicity, catalyst morphology, and specific facets or defect sites help stabilize intermediates and facilitate desorption [10–13]. Catalyst stability under operating potentials and electrolyte conditions is essential for sustained activity, resisting corrosion and deactivation. Using a highly conductive support with accessible porosity enables rapid electron and mass transfer, reduces overpotentials, and ensures uniform active-site utilization. Combining optimized intermediate binding, tailored electronic structure, stability, and nano-architecture forms the basis for effective two-electron electrocatalysts [8,14–17].

Cerium dioxide ($CeO_2$) is a promising electrocatalyst for $H_2O_2$ due to its unique properties, including redox cycling between $Ce^{3+}$ and $Ce^{4+}$, high surface oxygen vacancies, and stability. It enhances $O_2$ adsorption, activates $O_2$ at low overpotentials, and suppresses peroxide reduction to water. $CeO_2$, abundant, eco-friendly, and easily combined with supports or dopants, is efficient, cost-effective, and sustainable for $H_2O_2$ production efficiencies approaching 100% and hydrogen peroxide productivity in optimized systems [16,18–21].

In this sense, the present study aims to investigate the effect of metal loading on the ORR process. For this purpose, some scientific questions were considered: (i) whether the applied synthesis route could obtain $CeO_2$ nanowires. (ii) Can the MnOx surface modification improve the $H_2O_2$ electrosynthesis? (iii) How did the $CeO_2$ and $CeO_2$MnOx mass rate affect Vulcan XC-72 selectivity? (iv) Are synthesized nanoparticles suitable for i*n-situ* $H_2O_2$ electrogeneration applications?

## 2. MATERIALS AND METHODS

### 2.1 Reagents

All chemicals were of analytical grade and used without further purification. Potassium sulfate ($K_2SO_4$, ≥99%, CAS 7778-80-5), sulfuric acid ($H_2SO_4$, 98%, CAS 7664-93-9), cerium(III) nitrate hexahydrate ($Ce(NO_3)_3{\cdot}6H_2O$, ≥99%, CAS 10294-41-4), manganese(II) nitrate solution ($Mn(NO_3)_2$, ≥99%, CAS 10377-66-9), sodium hydroxide (NaOH, ≥98%, CAS 1310-73-2), nitric acid ($HNO_3$, ≥98%, CAS 1310-73-2), and hydrochloric acid (HCl, ≥98%, CAS 7647-01-0) were purchased from commercial suppliers. Aqueous solutions were prepared using ultrapure water (resistivity 18 $M\Omega{\cdot}cm^{-1}$ at 25°C) obtained from a Merck Millipore Milli-Q water purification system.

### 2.2 Synthesis of $CeO_2$ nanowires

$CeO_2$ nanowires were synthesized via a hydrothermal method[22,23]. Briefly, a 14.0 mol $L^{-1}$ NaOH solution was prepared and transferred into a 100 mL Teflon-lined stainless-steel autoclave. Subsequently, a 4.5 mol $L^{-1}$ $Ce(NO_3)_3$ $6H_2O$ solution was added dropwise while stirring. The autoclave was then sealed and heated at 110 °C for 24 h. After the reaction, the autoclave was allowed to cool naturally to room temperature. The resulting products were recovered by centrifugation and washed three times with ethanol and three times with deionized water to remove residual precursors. Finally, the as-obtained solids were dried at 85 °C for 24 h before further characterization [22].

#### 2.2.1 Surface-engineered $CeO_2MnOx$ heterostructure

10% $CeO_2MnOx$ composites were prepared by a hydrothermal deposition, where the 10% denotes the molar ratio of Mn to Ce. Briefly, a calculated amount of $CeO_2$ nanowires and the appropriate amount of $Mn(NO_3)_2$ were dissolved in 20 mL of deionized water; 38.4 g of NaOH was dissolved in 140 mL of deionized water. The two solutions were then combined under continuous stirring for 30 min to form a uniform suspension. The mixture was transferred into a 100 mL Teflon-lined stainless-steel autoclave and heated at 100 °C for 24 hours. The precipitate was subsequently collected by filtration, washed three times with ethanol and deionized water, and dried at 85 °C for 24 hours. Finally, the dried product was calcined at 450 °C for three hours to obtain the 10% $CeO_2$–$MnO_x$ composite [22,23].

#### 2.2.2 Electrocatalyst preparation

$CeO_2$ and $CeO_2MnO_x$ electrocatalysts were supported on Vulcan XC-72 at nominal loadings of 1, 3, and 5% and prepared by wet impregnation, following the procedure reported by Pinheiro et al. [20], which observed that 5 and 10 wt.% $CeO_2$ decreased electrocatalysts selectivity for $H_2O_2$ electrosynthesis. Briefly, the appropriate amount of $CeO_2$ and 10% $CeO_2$MnOx was added to 0.5 g of Vulcan XC72 and dispersed in 30 mL of deionized water under vigorous magnetic stirring for five hours. The resulting suspensions were then dried in an oven at 90 °C to obtain the supported electrocatalysts [18,21].

#### 2.2.3 Gas-diffusion electrode (GDE) preparation

Vulcan XC-72 and $CeO_2$MnOx electrocatalysts that showed the highest selectivity for $H_2O_2$ were applied for GDE preparation, based on mixing the electrocatalyst mass (Item 2.2.2.) with 20% w/w of Polytetrafluoroethylene (PTFE dispersion, 60% w/w in water, CAS 9002-84-0) in 500 mL of Milli-Q water under constant stirring for 24 hours. Subsequently, the electrocatalyst was vacuum-filtered and dried at 100 °C for 4 hours to achieve a dry mass of 2–10%. The obtained material was hot-pressed (SOLAB SL-10/15-E) between two stainless steel plates (3.5 $cm^2$) at 290°C under 4 tons for 2 hours [8,24].

## 2.3 Characterization

### 2.3.1 Physicochemical assessment

Transmission electron microscopy (TEM, JEOL JEM 2100) and Scanning electron microscopy (SEM, FESEM JEOL JSM-7401F) were employed to evaluate the morphology, aggregation state, and particle size of the synthesized $CeO_2$, 10% $CeO_2$–$MnO_x$ nanoparticles, and their incorporation in Vulcan XC72. Imaging was performed using a JEOL JEM-2100 microscope with an energy-dispersive X-ray spectroscopy (EDS) module for semi-quantitative elemental analysis. Samples were prepared on Formvar/carbon-coated copper grids (300 mesh, Merck) under a stereoscopic microscope (Digilab DI-152 T). The nanowire diameters were measured using ImageJ.

X-ray powder diffraction (XRD) patterns were acquired in transmission geometry using a STADI-P diffractometer (Stoe®, Darmstadt, Germany) operated at 40 kV and 40 mA with a Ge(111) primary-beam monochromator producing CuK$\alpha_1$ radiation ($\lambda$ = 1.54056 Å). Samples were mounted between two cellulose acetate sheets, and the holder was rotated continuously during data collection. Diffraction profiles were recorded over the 2$\theta$ range of 10–90 ° at a scan rate of 2 ° $min^{-1}$ using a Mythen 1 K linear detector (Dectris®, Baden, Switzerland).

X-ray photoelectron spectroscopy (XPS) was performed with a Scienta Omicron ESCA+ spectrometer using monochromatic Al K$\alpha$ radiation (1486.7 eV). Shirley's method was applied to correct the inelastic background of the C 1s spectra, and spectral deconvolution was performed using CasaXPS software with unconstrained, multiple-Voigt profile fitting [25].

Room-temperature Electron Paramagnetic Resonance (EPR) spectroscopy was carried out on a Bruker EMX Plus spectrometer. The experiments used an X-band (~9.4 GHz) resonator, with microwave power maintained at 10 mW. The acquired EPR spectra were subsequently simulated and fitted with derivative Lorentzian curves using the Microcal Origin program [26,27].

Wettability was determined by measuring water contact angles with a contact angle meter (SEO Phoenix 300, Kromtek). Briefly, 20 µL of nanoparticle suspension was deposited onto a glassy carbon substrate and dried, then a 10 µL droplet of deionized water was placed on the film surface. Images were captured and analyzed using Surfaceware9 software, and each measurement was conducted in triplicate.

#### 2.3.2 Electrochemical investigation

The electrocatalyst inks were prepared by dispersing 2 mg $mL^{-1}$ of the catalyst in deionized water under ultrasonic agitation and dropping 20 µL onto the glassy carbon disk. After drying under ambient conditions, 20 µL of a 1:100 (v/v) Nafion: deionized water solution was deposited onto the catalyst layer and dried thoroughly. Before measurement, the electrolyte was saturated with nitrogen for 30 min (electrochemical activation) and maintained under a continuous flow of oxygen during the measurements.

Rotating ring-disk electrode (RRDE) experiments were conducted at rotation rates of 100-1600 rpm using a potentiostat/galvanostat (Autolab PGSTAT 302 N). The working electrode consisted of a glassy carbon disk (0.2475 $cm^2$) surrounded by a platinum ring (0.1866 $cm^2$), yielding a collection efficiency (N) of 0.21[28]. A platinum plate and Ag/AgCl electrode were used as the counter and reference electrodes, respectively. The supporting electrolyte was 0.1 mol $L^{-1}$ $K_2SO_4$ (pH = 3.0).

Linear sweep voltammetry was carried out on the disk electrode in the potential range from 0.6 V to −0.9 V vs. Ag/AgCl at a negative scan rate, while the platinum ring was held at 1.0 V vs. Ag/AgCl to oxidize any $HO_2^-$ species reaching the ring to $O_2$. The number of electrons transferred in the ORR and the selectivity toward $H_2O_2$ production were determined using **Equations (1)** and **(2)**:

$$\%H_2O_2 = \frac{\frac{2I_r}{N}}{I_d + \frac{I_r}{N}} \times 100 \qquad (1)$$

$$n = 2\,[(X_{H2O2}) + 1] \qquad (2)$$

*Ir*, *Id*, and *N* represent the ring current, disk current, and RRDE collection efficiency, respectively [28].

2.3.3 *In-situ* $H_2O_2$ electrogeneration

The electrochemical experiments were conducted in an undivided cell containing the produced GDE (cathode) and platinum as an anode. Before starting electrosynthesis, the GDE was electroactivated with pressurized $N_2$ (0.2 bar) at 100 mA cm$^{-2}$ for 1 hour.

Subsequently, GDE was saturated with $O_2$ (0.2 bar) during $H_2O_2$ electrosynthesis, which was conducted with 350 mL of 0.1 mol L$^{-1}$ $K_2SO_4$ (pH 3.0) as the electrolyte and 137.5 mA cm$^{-2}$ for 1 hour, as reported previously [29]. The $H_2O_2$ yield was determined by spectrophotometry using the ammonium molybdate (2.4 mmol L$^{-1}$ with 0.5 mol L$^{-1}$ of $H_2SO_4$)[30], which contributed to determining the merit figures: specific production of $H_2O_2$ (kWh g$^{-1}$) **(Eq. 3)**, and Faradaic efficiency (CE%) **(Eq. 4)** [31–33].

$$SP_{H2O2}\,(kWh\,g^{-1}) = \frac{[H_2O_2] \times V}{E \times I \times t} \qquad (3)$$

$$FE(\%) = \frac{2\,F\,[H_2O_2]\,V}{I\,t} \times 100 \qquad (4)$$

Where 2 is the number of electrons exchanged for the reduction of $O_2$ to $H_2O_2$, F is the Faraday constant (96485 C mol$^{-1}$), [$H_2O_2$] is the hydrogen peroxide concentration mol L$^{-1}$ (**Eq. 4**), and g L$^{-1}$ (**Eq. 3**). V is the working volume (L), I is the applied electric current (A), E is the cell potential (V), and t is the electrolysis time (s) (**Eq. 3**) and h (**Eq. 4**).

# 3. RESULTS AND DISCUSSION

## 3.1 $CeO_2$ and $CeO_2MnOx$ characterization

The morphologies of materials were investigated by TEM and FE-SEM micrographs. TEM image of the $CeO_2$ material, illustrated in **Figure 1a**, revealed the formation of the expected anisotropic one-dimensional morphology, with well-defined and grouped nanowires. An EDS spectrum obtained in the region of one of the wires (**Figure 1b)** detected the presence of Ce and O as the constituent elements of the composition, as expected. The histogram of the wire diameter distribution, constructed from the counting of 100 nanowires, indicated diameters ranging from 5.1 to 15.7 nm. These results suggest that the established synthesis route is an efficient surfactant-free strategy for nanowire formation.

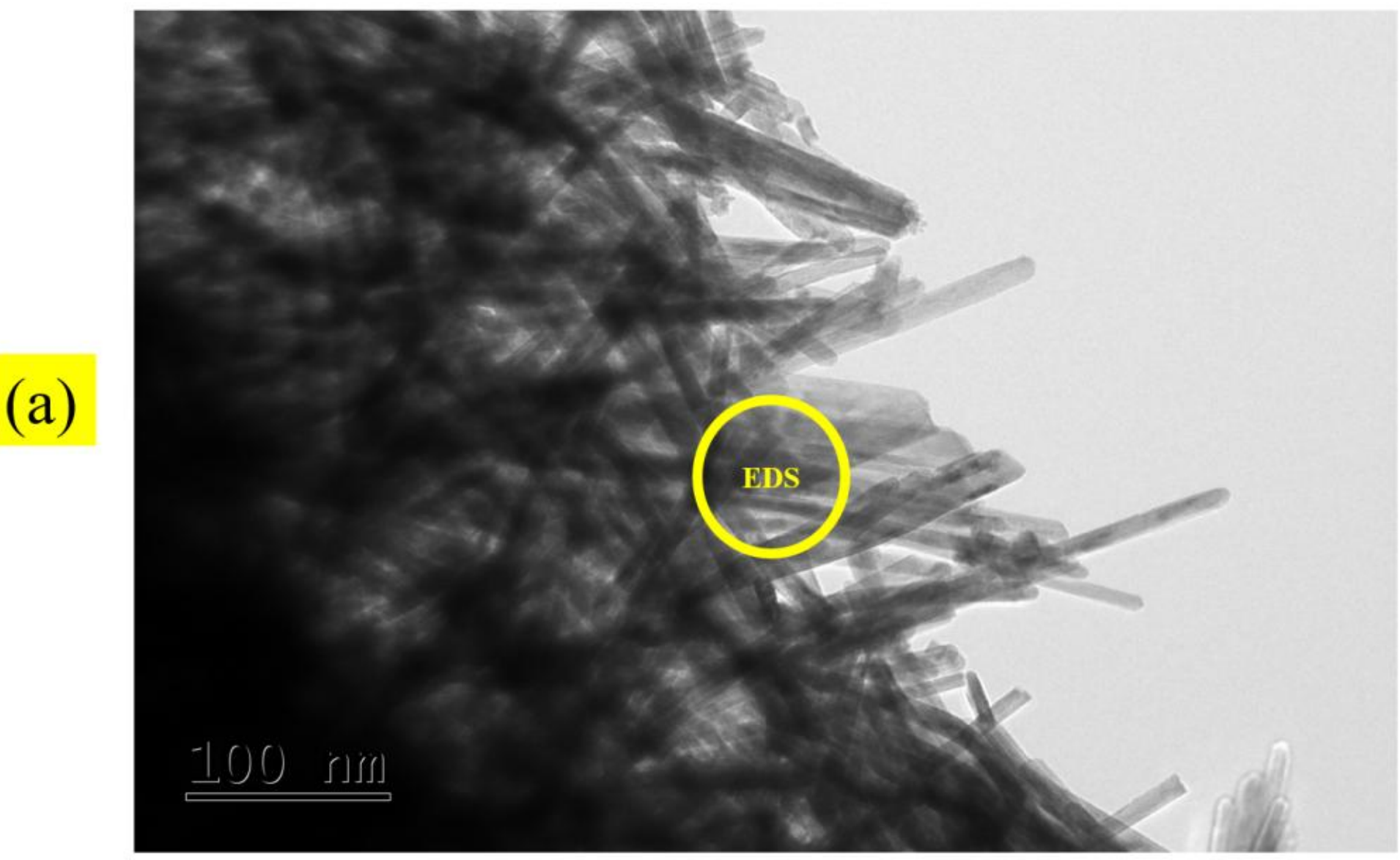

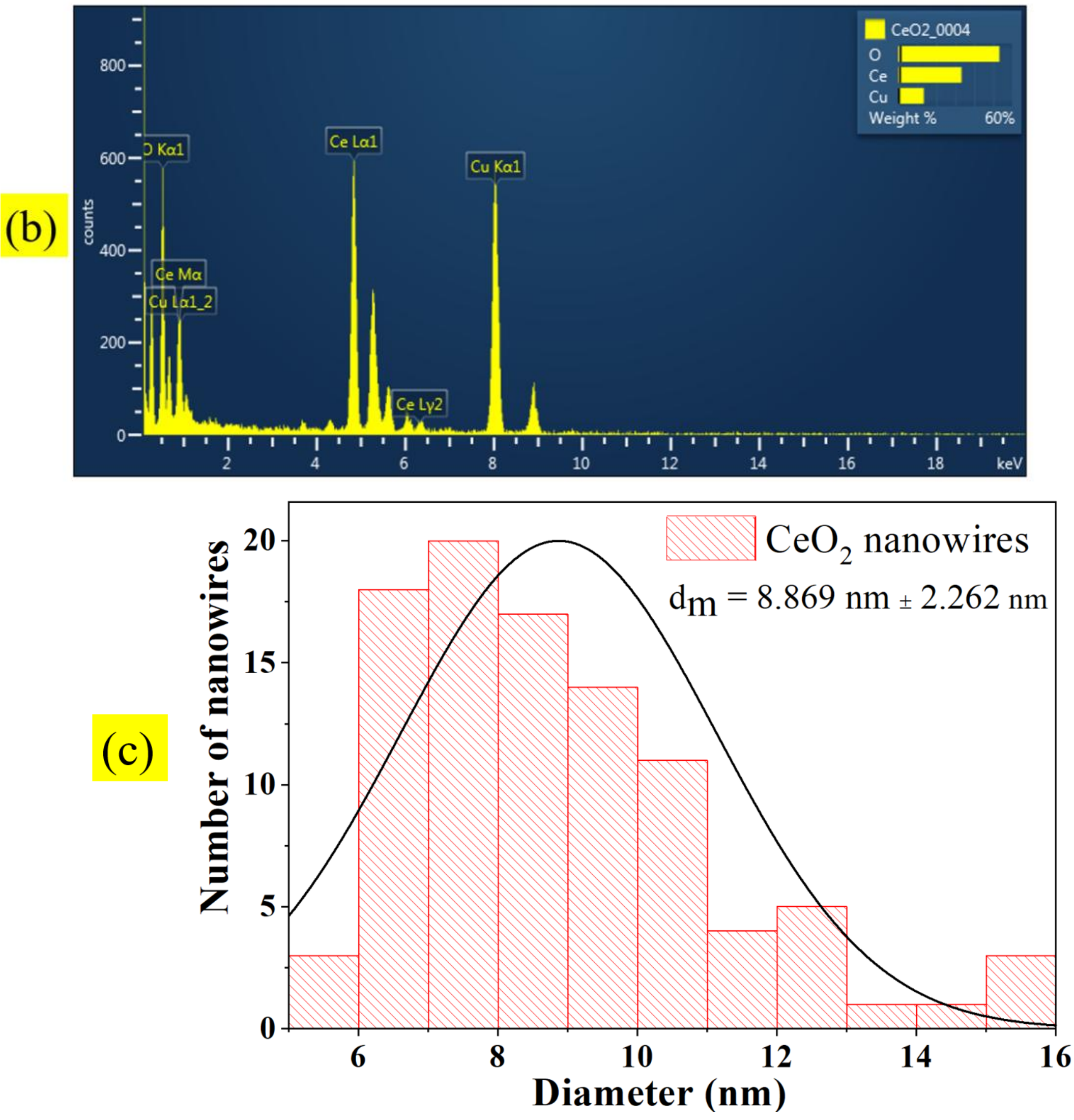

**Figure 1.** (a) TEM image and (b) EDS spectra of the pure $CeO_2$ nanowires, (c) histogram of the distribution of the average diameter size of the nanowires ($d_m$).

TEM and FE-SEM images obtained for the $CeO_2MnOx$ heterostructure are illustrated in **Figure 2a** and **Figure 2b**. It is possible to observe nanowires and particulate clusters around and on the nanowires. EDS spectra (**inset, Figure 2b)** obtained in different regions of the sample detected Mn in the particulate region of the structure and predominantly Ce in the elongated region (**Figure 2b**). These results indicate the formation of a $CeO_2MnOx$ heterostructure, as observed by Ahmad et al.[34].

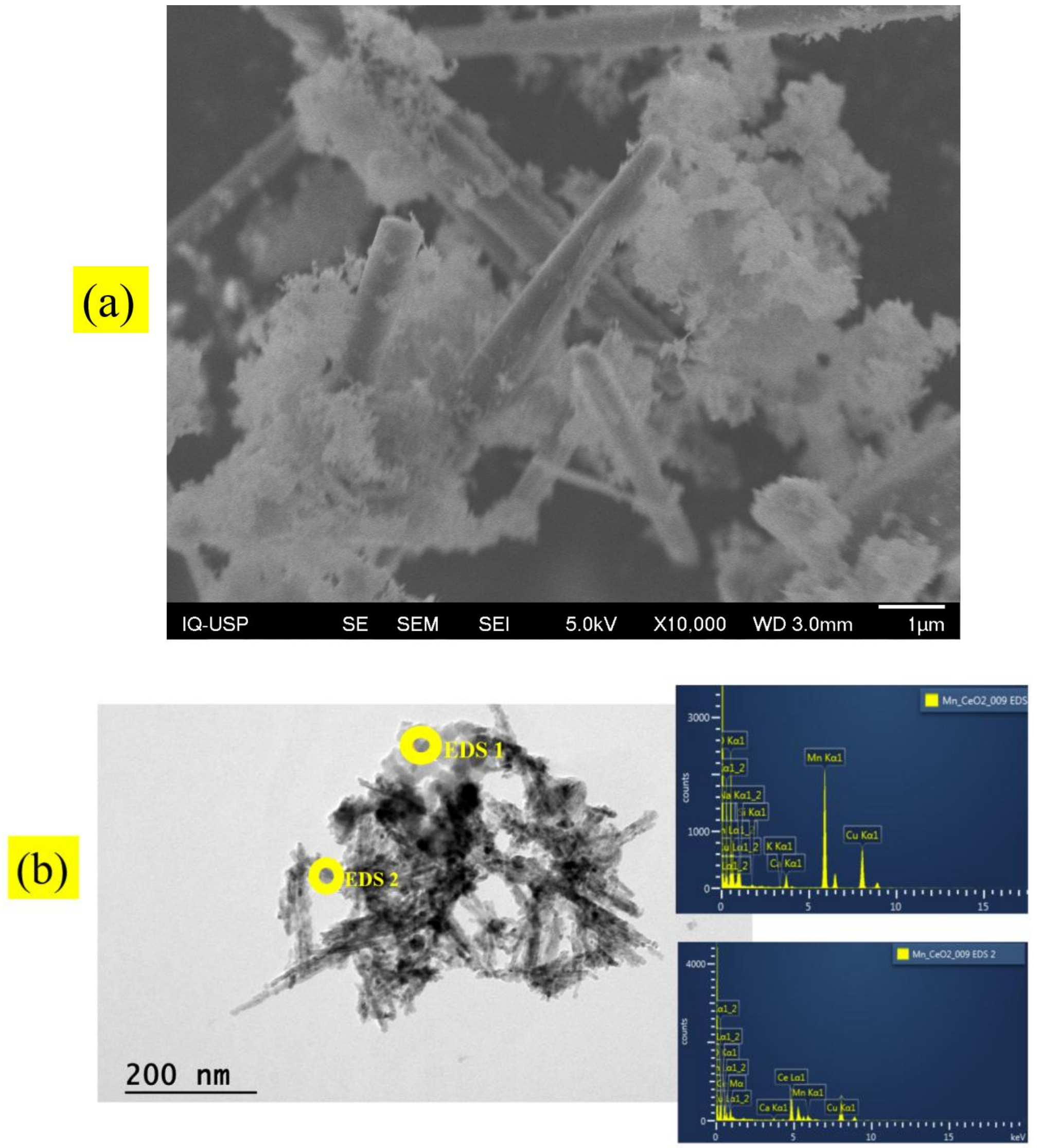

**Figure 2.** (a) TEM and (b) FE-SEM images of the $CeO_2MnOx$ heterostructure and inset (b) EDS spectra.

**Figure 3** shows the XRD patterns of $CeO_2$ and $CeO_2MnOx$ heterostructure. The diffraction peaks observed for the pure $CeO_2$ sample can be indexed to the face-centered cubic fluorite structure of $CeO_2$ (Fm3m), as according to Inorganic Crystal Structure Database (ICSD) card No. 182182, the prominent reflections were detected at approximately $2\theta$ = 28.5°, 33.1°, 47.5°, 56.3°, 59.1°, 69.4°, 76.7°, and 79.1°, corresponding to the (111), (200), (220), (311), (222), (400), (331), and (420) lattice planes, respectively. No secondary phases were detected, indicating the high crystallinity and phase purity of the as-synthesized $CeO_2$. Upon MnOx addition (**Figure 3**), the diffraction peaks of the $CeO_2$ phase are preserved. In contrast, an additional peak was associated with the hausmannite $Mn_3O_4$ oxide phase, as 2θ = 38.2° corresponding to the $I4_1/amd$ space group (ICSD No. 33327)[35]. All the presented evidence suggests that $CeO_2$ and $CeO_2MnOx$ were successfully synthesized and modified, yielding improved surfaces that enhanced ORR.

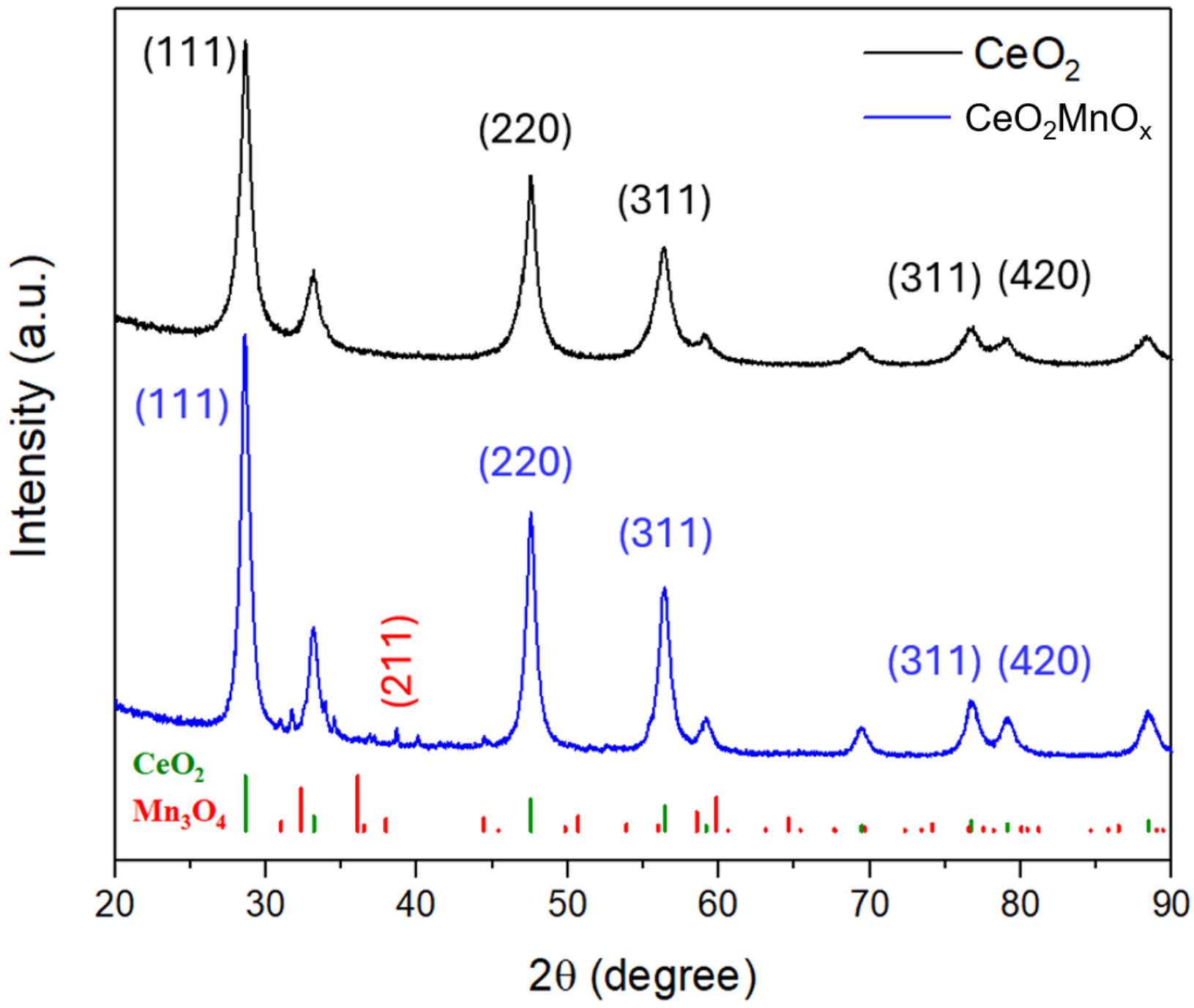

**Figure 3**. XRD patterns of synthesized $CeO_2$ and $CeO_2MnOx$.

XPS analysis was conducted on $CeO_2$ and $CeO_2MnOx$ samples to verify the chemical composition of the synthesized pure materials (**Figure 4**). **Figure 4a** shows the Ce 3d spectrum of pure $CeO_2$. The spectrum displays peaks for $Ce^{3+}$ 3d5/2 and 3d3/2 at 882 and 900 eV, respectively, along with peaks for $Ce^{4+}$ 3d5/2 and 3d3/2 at 897 and 916 eV, respectively. Satellite peaks indicative of $Ce^{3+}$ were also detected. In the $CeO_2MnOx$ sample (**Figure 4b)**, the Ce 3d spectrum shows an increased relative intensity of $Ce^{3+}$ components ($v_0$, v', $u_0$, u') compared to pure $CeO_2$. Since $Ce^{3+}$ formation involves charge compensation via oxygen vacancies, this suggests a higher concentration of oxygen defects. The Mn 2p spectrum further reveals the presence of both $Mn^{3+}$ and $Mn^{4+}$, with $Mn^{3+}$ known to facilitate oxygen removal from the lattice (**Figure 4c**). These combined Ce 3d and Mn 2p results confirm that manganese incorporation affects the material's defect chemistry, enhancing the formation of oxygen vacancies in $CeO_2$ [34,36].

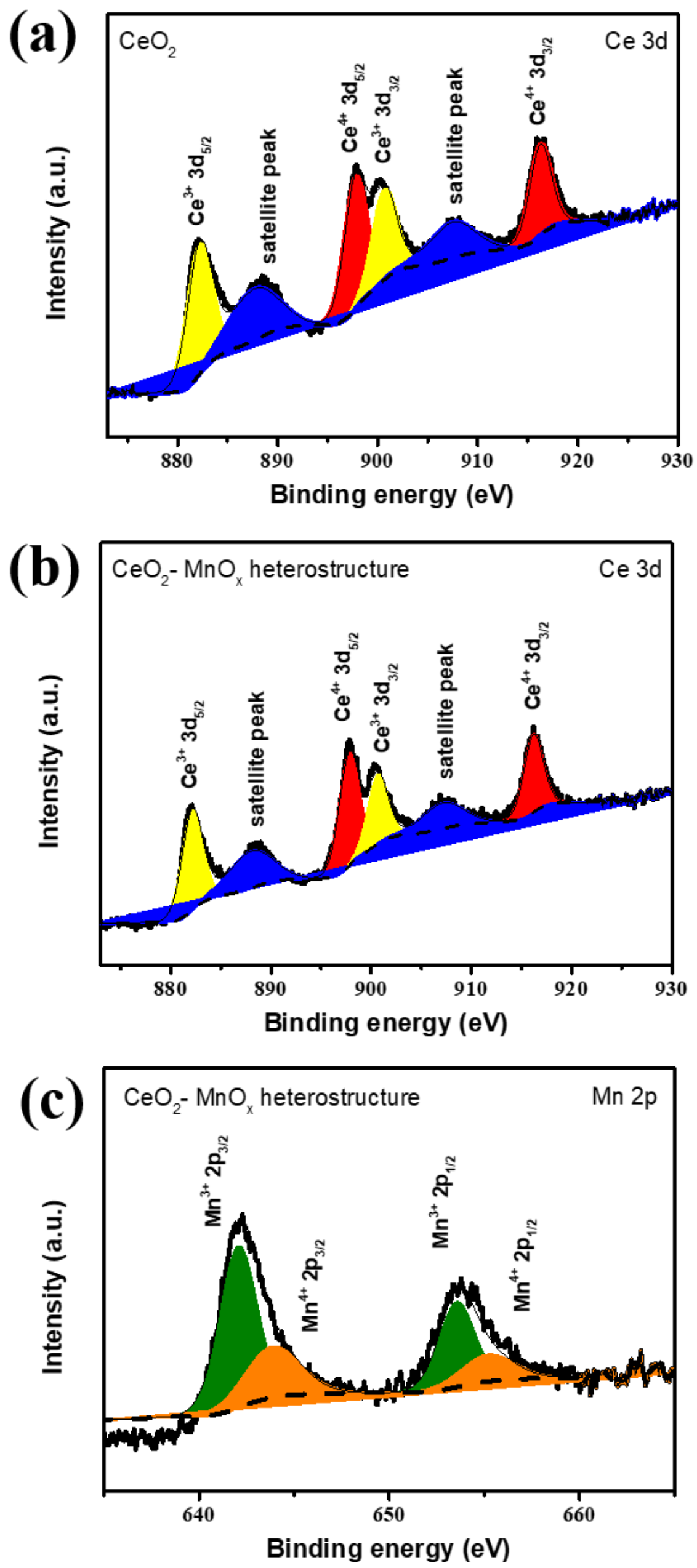

**Figure 4.** High-resolution spectra of Ce 3d for (a) $CeO_2$, (b) $CeO_2MnOx$, and (c) high-resolution spectra of Mn 2p for $CeO_2MnOx$.

**Figure 5** presents XPS analyses of the Vulcan XC72 carbon, 3% $CeO_2$/C, and 1% $CeO_2MnOx$/C electrocatalysts, performed to investigate and quantify the functional groups present. **Figure 5** was deconvoluted into contributions from different functional groups on the carbon atom. The functional groups identified were C-C, C-OH, C=O, and -COOH, located at approximately 284, 285, 286, and 289 eV, respectively. It can be noted that the percentage of oxygenated species present in the 3% $CeO_2$/C sample is higher than that of the Vulcan XC-72 sample, indicating that the modification of the carbon with $CeO_2$ has altered the carbon surface. The 1% $CeO_2MnOx$/C sample shows the highest percentage of oxygenated species among the samples, indicating that the synergistic effect of $CeO_2$ and Mn has further altered the carbon surface [37].

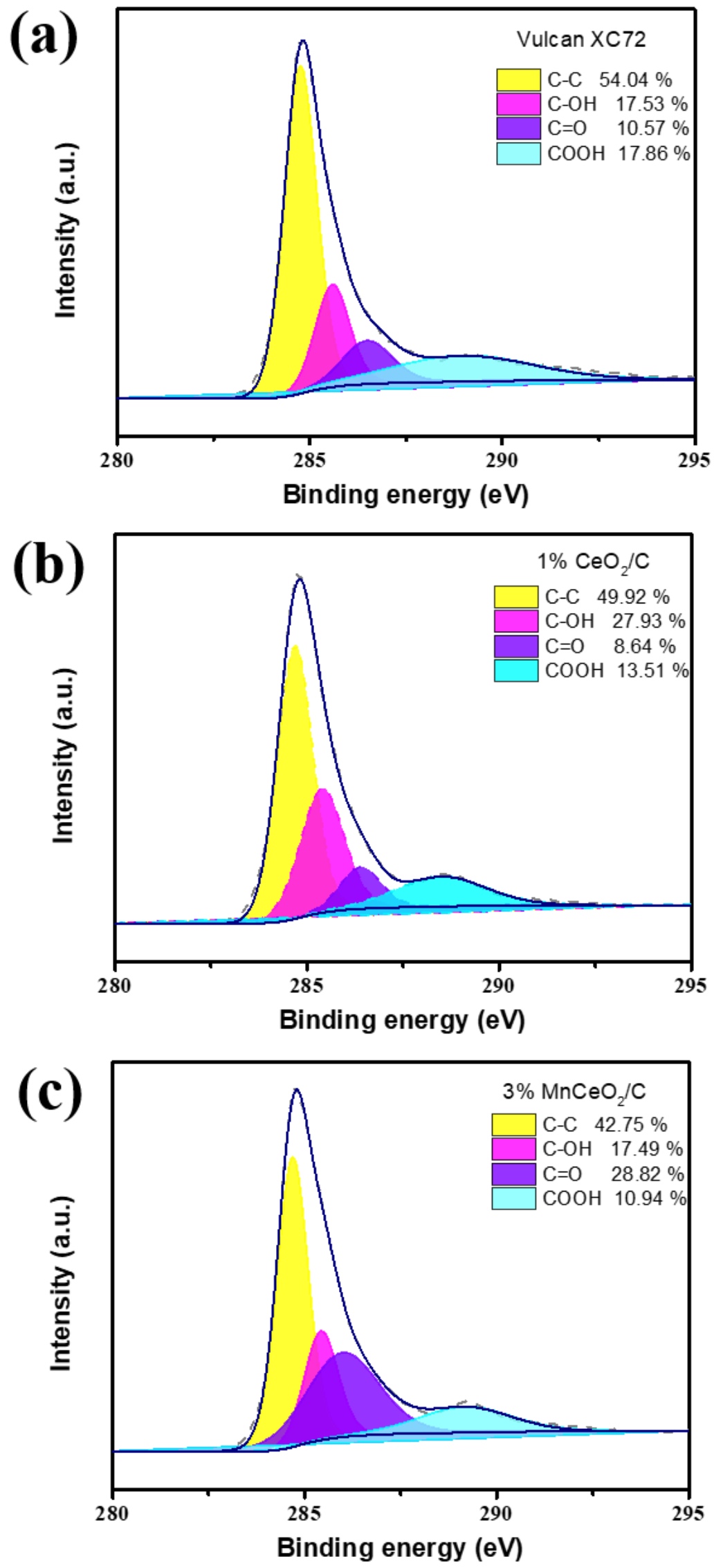

**Figure 5.** High-resolution spectra of C 1s for (a) Vulcan XC-72, (b) 3% $CeO_2$/C, and (c) 1% $CeO_2MnOx$/C.

In summary, adding cerium oxide to a carbonaceous support is well known for boosting local oxygen levels, thanks to its remarkable ability to store and release oxygen [18]. Its effectiveness stems from a combination of structural, kinetic, and textural traits, such as its capacity to switch between $Ce^{4+}$ and $Ce^{3+}$ oxidation states reversibly, and to form both stoichiometric and non-stoichiometric phases ($CeO_2$ and $CeO_{2-x}$).

$$CeO_2 \leftrightarrow CeO_{2-X} + \left(\frac{X}{2}\right) O_2 \; (O \le x \ge 0.5) \qquad (5)$$

This reversible reaction (**Eq. 5**) exemplifies ceria's role as a versatile oxygen reservoir, providing molecular oxygen under reducing conditions. Notably, oxygen vacancies in $CeO_{2-x}$ enhance oxygen mobility relative to fully stoichiometric cerium oxide, facilitating more rapid diffusion to active sites. Accordingly, integrating ceria into electrocatalysts enables exploitation of its defect chemistry to maintain elevated local oxygen pressures and enhance electrochemical processes, including peroxide formation [18].

**Figure 6** presents the EPR spectra of the pure $CeO_2$ and $CeO_2MnO_x$ catalysts, which provide critical insight into the role of oxygen vacancies. The pure $CeO_2$ sample exhibits a broad resonance signal characteristic of ferromagnetic interactions. In ceria-based nanostructures, such behavior is frequently associated with the presence of oxygen vacancies, which can induce localized magnetic moments and facilitate ferromagnetic coupling[26,27].

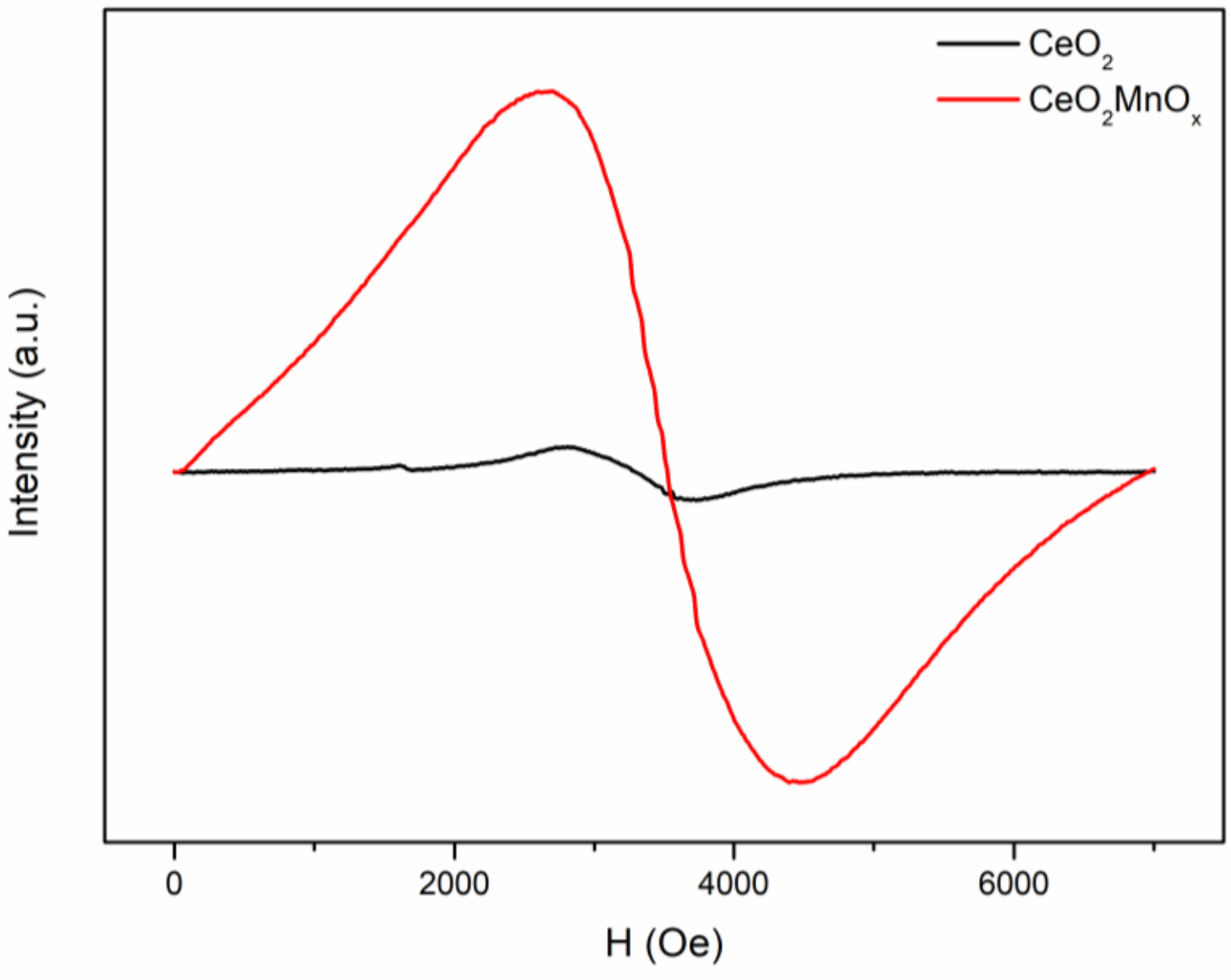

**Figure 6.** EPR spectrum of $CeO_2$ and $CeO_2MnO_x$ at room temperature.

A profound enhancement of this signal is observed upon incorporation of 15% Mn, as shown in **Figure 6**. Quantitative integration of the EPR signal areas reveals a 30-fold increase for the $CeO_2MnO_x$ sample compared to the unmodified $CeO_2$. This dramatic signal amplification cannot be rationalized solely by the ferromagnetic contribution from the added $MnO_x$. If the signal were primarily due to $MnO_x$, the introduction of a 15 wt.% Mn loading would be expected to yield a significantly smaller increase. The observed 30-fold enhancement far exceeds this expectation, strongly indicating that the Mn incorporation does not merely add ferromagnetic contribution from the oxide but actively promotes the generation of a high concentration of additional oxygen vacancies within the $CeO_2$ lattice. This synergistic effect creates a defect-rich nanostructure, which we posit is a key factor in enhancing the $H_2O_2$ electrosynthesis activity [38].

**Figure 7** presents the contact angle values of Vulcan XC-72, $CeO_2$/C, and $CeO_2MnOx$/C electrocatalysts, including their respective statistical groupings according to Tukey's test ($p < 0.05$). Pristine Vulcan exhibited a contact angle of approximately 77.6° (group b), consistent with its moderately hydrophobic surface, and the incorporation of 1 wt.% $CeO_2$/C led to a marked reduction in contact angle (54.1°; group a), evidencing a significant increase in surface polarity relative to all other materials. This pronounced hydrophilicity can be attributed to the introduction of oxygenated surface species (e.g., Ce–OH, $Ce^{3+}$-associated vacancies), which enhance water adsorption and spreading.

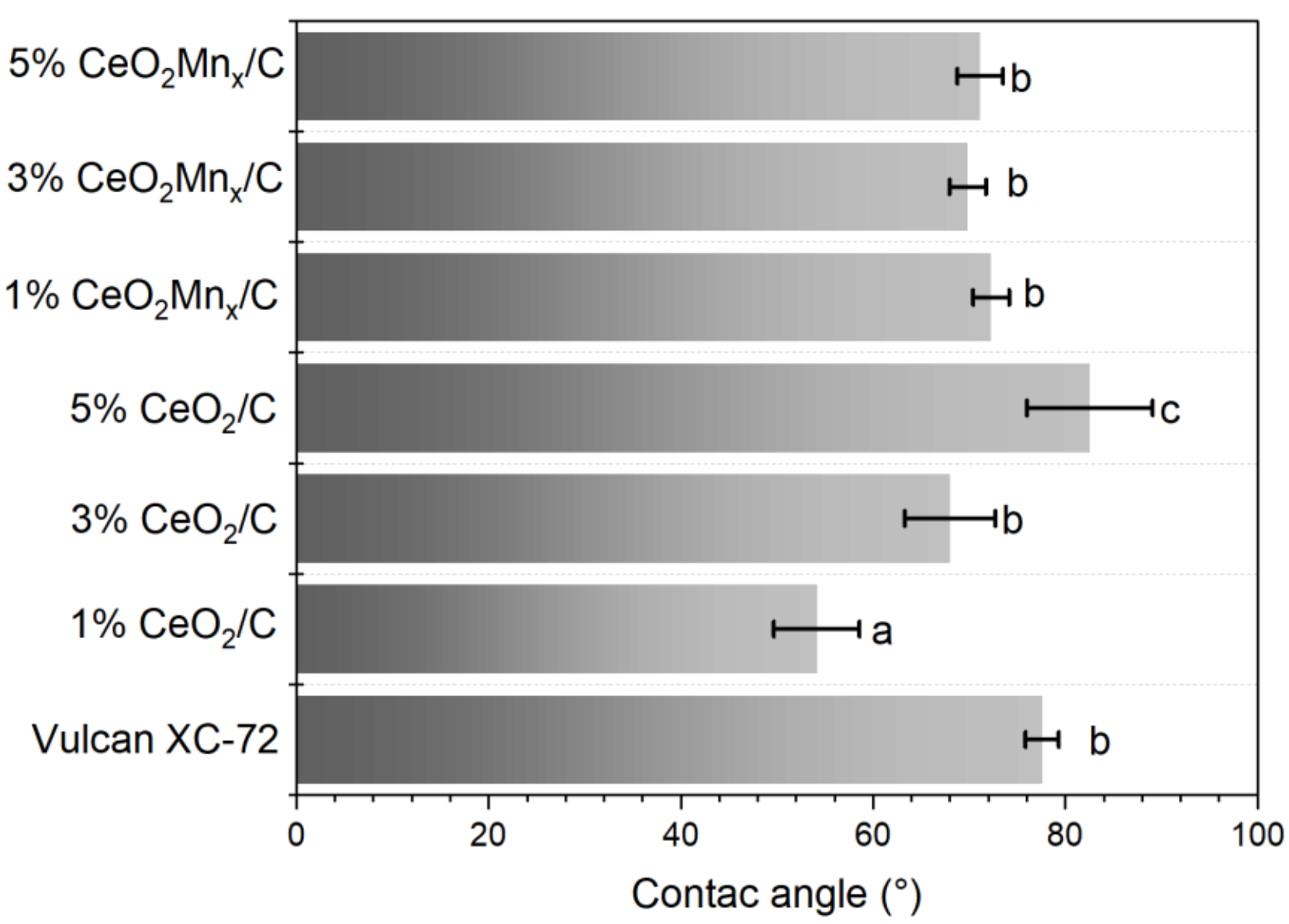

**Figure 7.** Average contact angle of Vulcan XC72, $CeO_2/C$, and $CeO_2MnOx/C$ electrocatalysts. (n=3). Different letters indicate statistically significant differences between groups according to Tukey's test ($p < 0.05$).

In contrast, the 3% $CeO_2/C$ sample showed a substantial increase in contact angle (68.0°; group b), which was statistically indistinguishable from that of Vulcan and the $MnCeO_2$ series. This reversal in wettability likely reflects the onset of $CeO_2$ aggregation at higher loadings, which can decrease the accessible polar surface and hinder uniform water spreading. Meanwhile, the 5% $CeO_2/C$ material exhibited the lowest contact angle among the $CeO_2$ samples (82.5°; group c), significantly higher than those of all others except Vulcan, suggesting a shift toward a more hydrophobic regime. This behavior is often associated with surface roughness and the possible coverage of hydrophilic carbon domains by larger $CeO_2$ clusters, which create patches of poor wettability.

Mn incorporation promoted a markedly different trend; the $CeO_2MnOx/C$ electrocatalysts, in all metal loadings, showed intermediate contact angle values (~70–72°; all in group b), forming a statistically homogeneous cluster. This behavior suggests that Mn addition introduces defect-rich structures (Mn–O–Ce linkages, oxygen vacancy modulation) that stabilize moderately hydrophilic surfaces across compositions. These features can facilitate water adsorption and electrochemical wetting, supporting efficient electrolyte penetration and interfacial charge transfer [39].

### 3.2 Effect of $CeO_2$/C and $CeO_2MnO_x$/C on ORR

**Figure 8a** presents the RRDE polarization curves for a series of catalysts comprising Vulcan XC72 and different $CeO_2$/C and $CeO_2MnO_x$/C in various proportions, measured in $O_2$-saturated electrolyte at a scan rate of 5 mV $s^{-1}$. The disk currents ($I_{DISK}$) in the cathodic region indicate ORR activity. In contrast, the corresponding ring currents ($I_{RING}$) reflect the amount of $H_2O_2$ produced during ORR as detected at the ring electrode.

Vulcan XC-72 shows the most negative onset potential and the lowest ring current, demonstrating its baseline performance with a mixed two- and four-electron pathway **(Figure 8b)**. In contrast, catalysts containing $CeO_2$ and $CeO_2MnO_x$ nanostructures on carbon show significantly higher ring currents and increased disk currents. At less negative potentials, the 1% $CeO_2MnO_x$/C and 3% $CeO_2$/C electrocatalysts delivered the most pronounced activity toward the selective two-electron ORR pathway (**Figure 8a**), 90 to 70% **(Figure 8b)**, respectively. Specifically, these electrocatalysts yield higher peroxide levels, as indicated by greater ring current intensities (~50 μA) **(Figure 8a)**, suggesting enhanced peroxide electrogeneration efficiency.

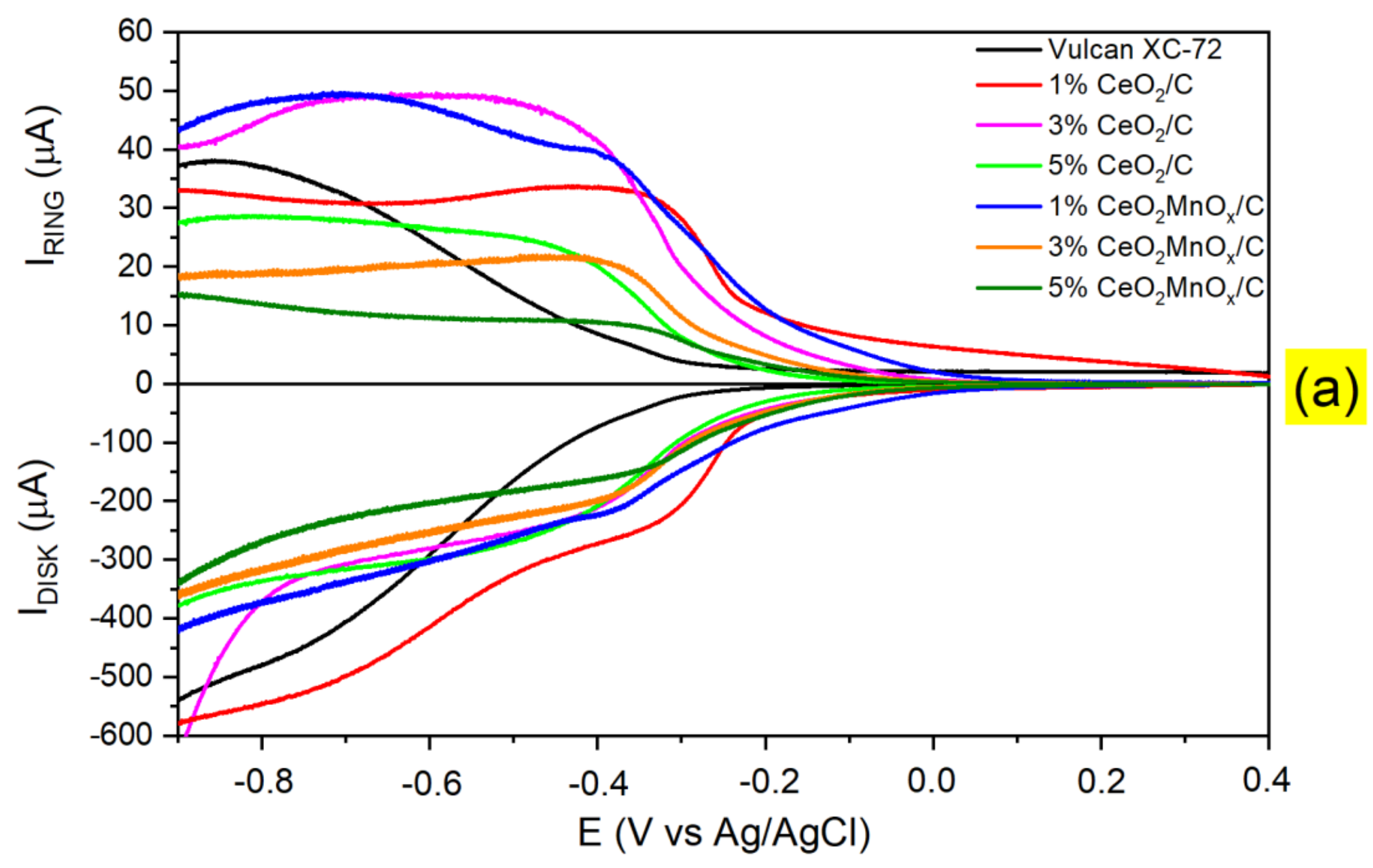

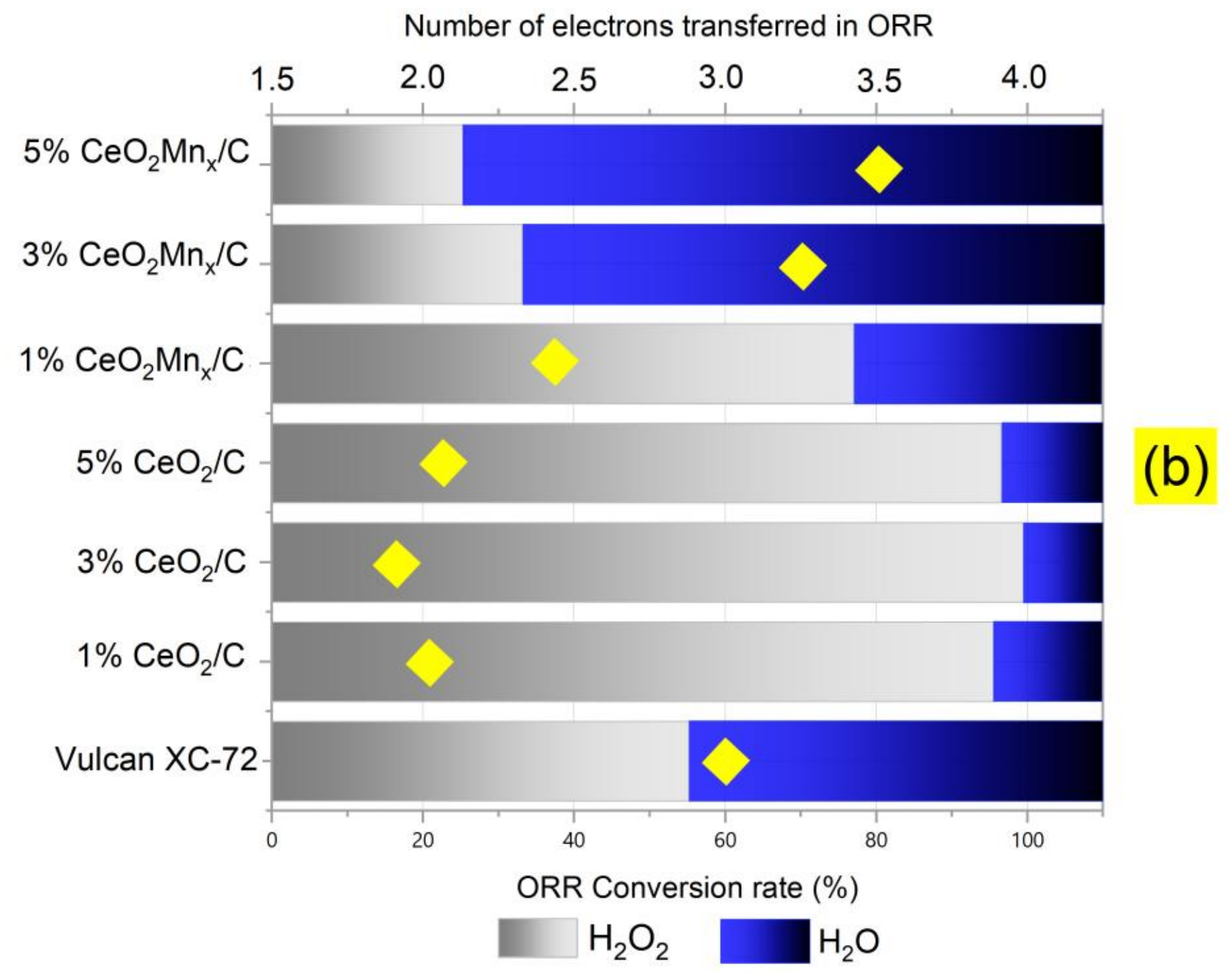

**Figure 8.** (a) Steady state polarization curves for ORR on different proportions of $CeO_2$ and $CeO_2MnOx$ on Vulcan XC-72. (b) Number of transferred electrons and $O_2$ conversion rate on the electrocatalysts. Experimental conditions: 0.1 mol $L^{-1}$ $K_2SO_4$, 0.01

mol $L^{-1}$ $H_2SO_4$ (pH =3.0) at a scan rate: 5 mV $s^{-1}$; ring current $E_{ring}$ = -0.6 V; disk current at 900 rpm.

**Table 1** presents a brief overview of the literature, highlighting the use of CeO2 and MnOx in electrocatalysis, to emphasize the goal and results obtained in the present study regarding the $H_2O_2$ electrocatalyst selectivity and the number of electrons transferred in ORR.

**Table 1**. Brief overview of $CeO_2$ and MnOx investigated for *in-situ* $H_2O_2$ electrogeneration.

| Nanostructures | Nt* | $H_2O_2$ (%) | Reference |
|---|---|---|---|
| 3% $CeO_2$/C | 1.87 | 99.5 | This work |
| 1% $CeO_2$Mn/C | 2.46 | 77.0 | This work |
| 1% $CeO_2$HARN/C | 2.1 | 95.0 | [20] |
| 4% $CeO_2$/C | 3.1 | 44.0 | [19] |
| 4% $CeO_2$ nanorods/C | 2.18 | 91.0 | [6] |

*Nt = number of electrons transferred in ORR

Assumpção *et al.* [40] investigated the effects of different preparation methods and carbon supports on the electrogeneration of hydrogen peroxide ($H_2O_2$) via ORR catalyzed by cerium oxide nanoparticles ($CeO_2$/C). The electrocatalysts were synthesized at a fixed 4% $CeO_2$ loading using two distinct methods — polymeric precursor (PPM) and sol-gel (SGM) and supported on Vulcan XC-72R or Printex L6 carbons. Electrochemical measurements using rotating ring-disk electrodes in alkaline media demonstrated that this catalyst exhibited the most efficient two-electron ORR pathway for $H_2O_2$ production, achieving ~88% selectivity and a low onset potential (-0.10 V). Hence, the combination of PPM synthesis and Printex L6 support yielded an optimal electrocatalyst with an increased surface oxygen content and enhanced catalytic activity for the selective generation of $H_2O_2$.

As noted by Assumpção et al. [40] ceria nanoparticles contribute to H2O2 electrogeneration, whereas using the lowest metal charge in the electrocatalyst yields similar results. The addition of Mn on the $CeO_2$ structure, without surfactant additions, provided an enhancement of 20% for $H_2O_2$ selectivity at 1% $CeO_2Mn/C$ in comparison to Vulcan XC-72 [40].

The kinetic parameters of the ORR were analyzed using the Koutecky–Levich (K–L plots) based on **Equations 6 and 7**, which separates the measured current density ($j$) into kinetic and diffusional contributions according to $j^{-1} = j_k^{-1} + (B\,\omega^{1/2})^{-1}$, where $j_k$is the kinetic current density and $\omega$is the angular rotation rate of the RRDE. B, Levich constant (**Eq. 7**) represents the mass-transport capability of the system, being determined by the number of electrons transferred per $O_2$ molecule (n), Faraday constant (F = 96485 C $mol^{-1}$), $C_{O2}$ the bulk oxygen concentration ($1.2 \times 10^{-6}$ mol $cm^{-3}$), $D_{O2}$ the bulk oxygen diffusion coefficient ($1.9 \times 10^{-5}$ $cm^2$ $s^{-1}$), ω is the angular rotation speed (rad $s^{-1}$), and V the kinematic electrolyte viscosity (0.01 $cm^2$ $s^{-1}$).

$$\frac{1}{j} = \frac{1}{j_k} + \frac{1}{j_d} = \frac{1}{j_k} + \frac{1}{B\omega^{1/2}} \qquad (6)$$

$$B = 0.62nFC_{O_2}D_{O_2}^{2/3}V^{-1/6} \qquad (7)$$

**Figure 9** illustrates the Koutecky–Levich plots for Vulcan XC-72, $CeO_2/C$, and $CeO_2MnOx/C$ electrocatalysts, clearly demonstrating that the metal loading significantly influences the ORR electroactivity.

The introduction of $CeO_2$ (**Figure 9a**) shifts the K–L lines to lower $j^{-1}$ values than those of bare Vulcan, indicating higher current densities and improved ORR kinetics. Among the ceria samples, 3% $CeO_2/C$ exhibits the lowest values of $j^{-1}$ over the whole range and the smallest intercept ($65.19\omega^{-1/2}$), which can be associated with the highest kinetic current density (3.7 mA cm$^{-2}$). In contrast, 5% $CeO_2$/C shows a partial loss of activity relative to 3%, suggesting that excessive oxide loading leads to particle agglomeration and/or blockage of the conductive carbon surface, limiting oxygen transport and the number of electroactive sites [16].

Similar slopes for Vulcan and 5% $CeO_2$/C (142.33 and $142.35\omega^{-1/2}$) indicate that the average number of electrons involved in the ORR does not change drastically with $CeO_2$ content and remains close to the 2-electron regime **(Figure 9b)**. It only exerts a moderate effect on the number of electrons involved and, consequently, the selectivity towards the two-electron pathway [16,20].

**Figure 9a** highlights this behavior through the curve's slope, indicating a first-order dependence on $O_2$, in the following order: Vulcan CX-72 > 5% $CeO_2$/C > 1% $CeO_2$/C > 3% $CeO_2$/C. Such modifications tend to facilitate mass transfer of dissolved oxygen to the electrocatalyst surface, thereby enhancing peroxide electrogeneration and catalytic activity, consistent with previous studies [16,20].

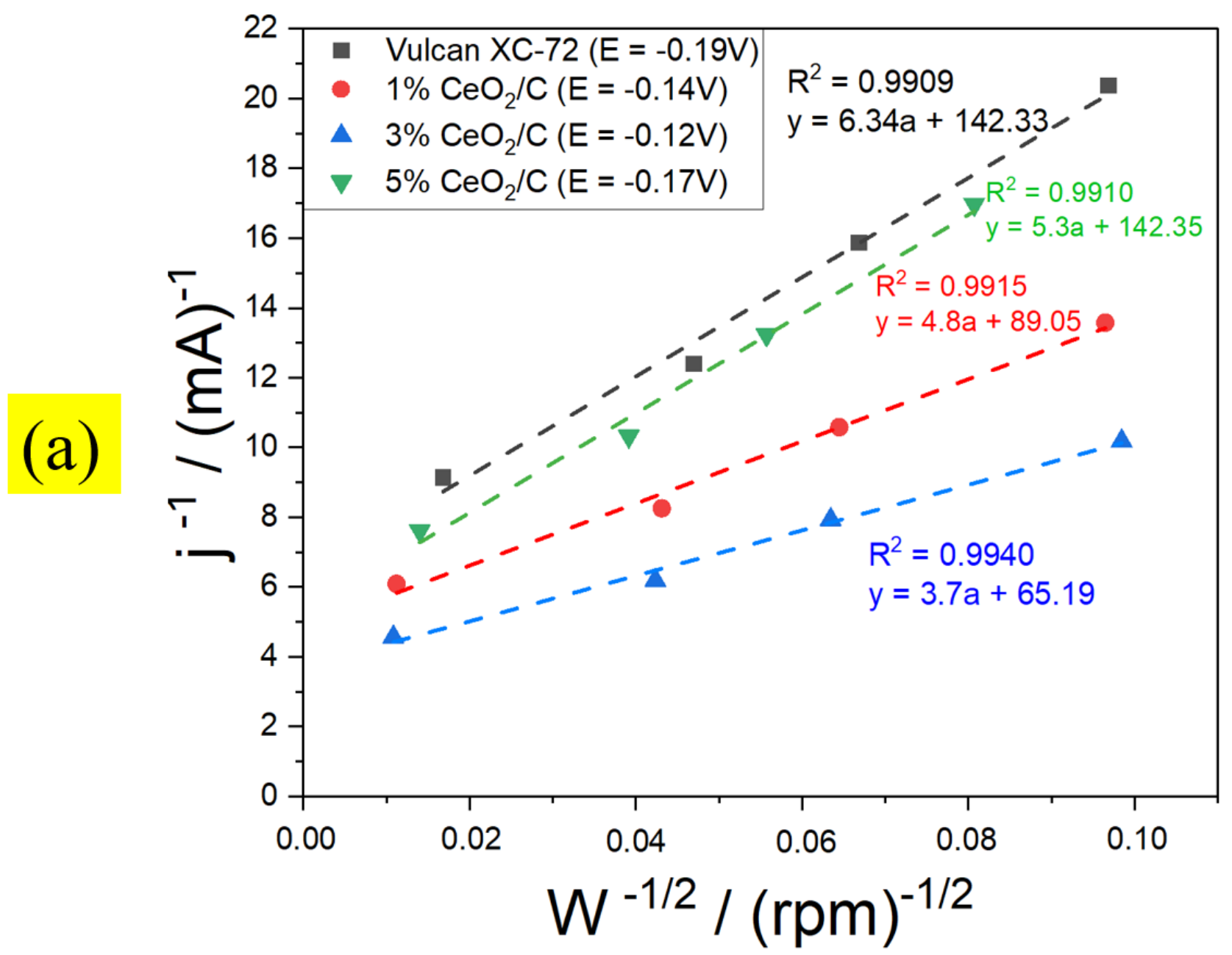

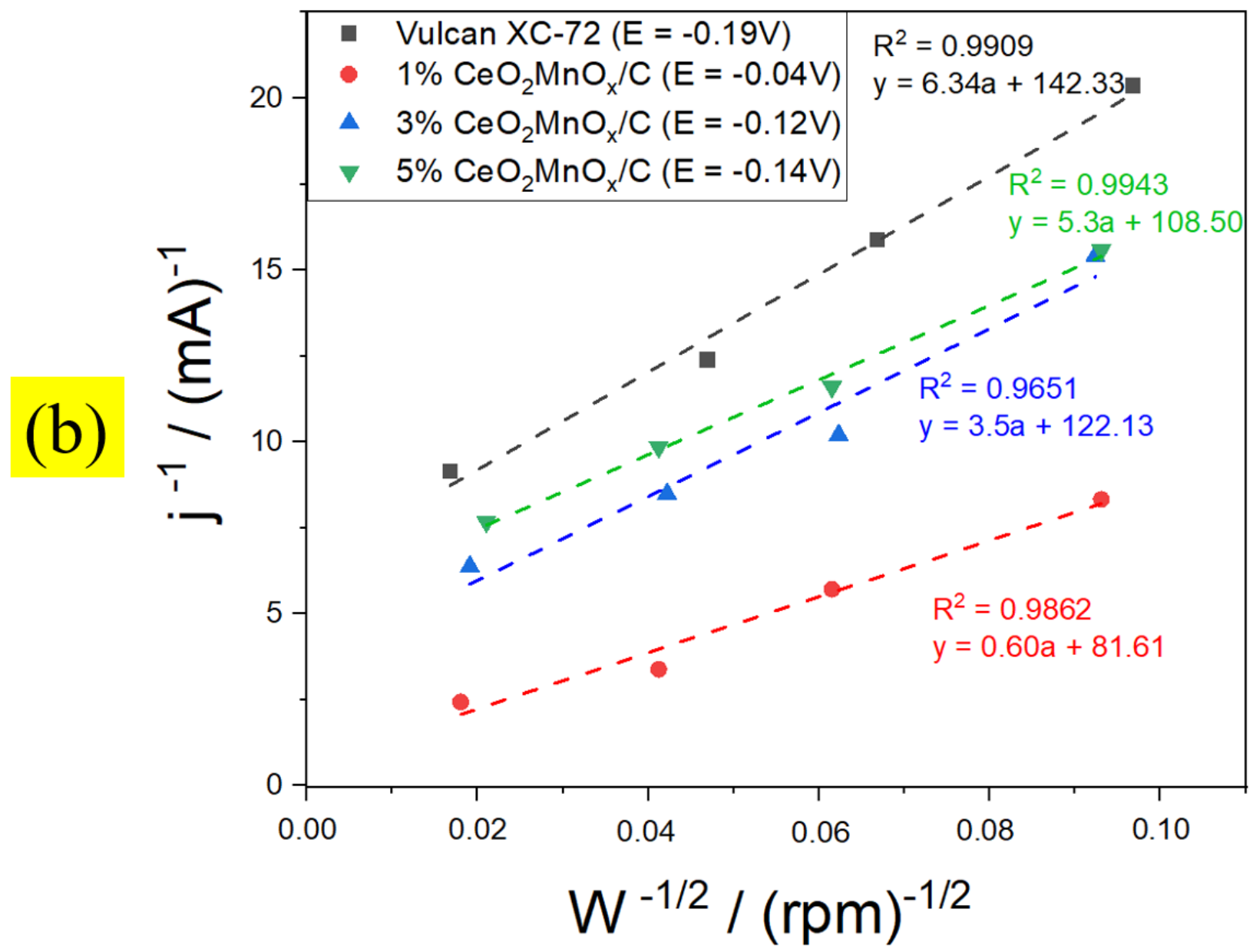

**Figure 9.** Koutecky–Levich plot for ORR on (a) $CeO_2$/C and (b) $CeO_2MnOx$/C on Vulcan XC-72. Experimental conditions: 0.1 mol $L^{-1}$ $K_2SO_4$, 0.01 mol $L^{-1}$ $H_2SO_4$ (pH =3.0) at a scan rate: 5 mV $s^{-1}$; ring current $E_{ring}$ = -0.6 V; disk current at 1600 rpm.

The addition of manganese oxide (**Figure 9b**) rendered the electrocatalyst surface more hydrophilic, as evidenced by the contact angle data (**Figure 7**), thereby increasing the proportion of oxygenated species, as indicated by the XPS data (**Figure 4**). Regarding $CeO_2MnOx$/C samples and their curve's slope, indicating a first-order dependence on $O_2$, follows: Vulcan CX-72 > 3% $CeO_2MnOx$/C > 5% $CeO_2MnOx$/C > 1% $CeO_2MnOx$/C.

### 3.3 *In-situ* $H_2O_2$ electrosynthesis assessment

Based on the previous results, the $CeO_2MnOx$ electrocatalyst that showed high selectivity for $H_2O_2$ and two-electron transfer in RRO was 1% $CeO_2MnOx$/C. In this sense, the electrocatalyst was used to prepare GDEs and compared with Vulcan XC-72 in an electrolysis for 60 min (**Figure 10**).

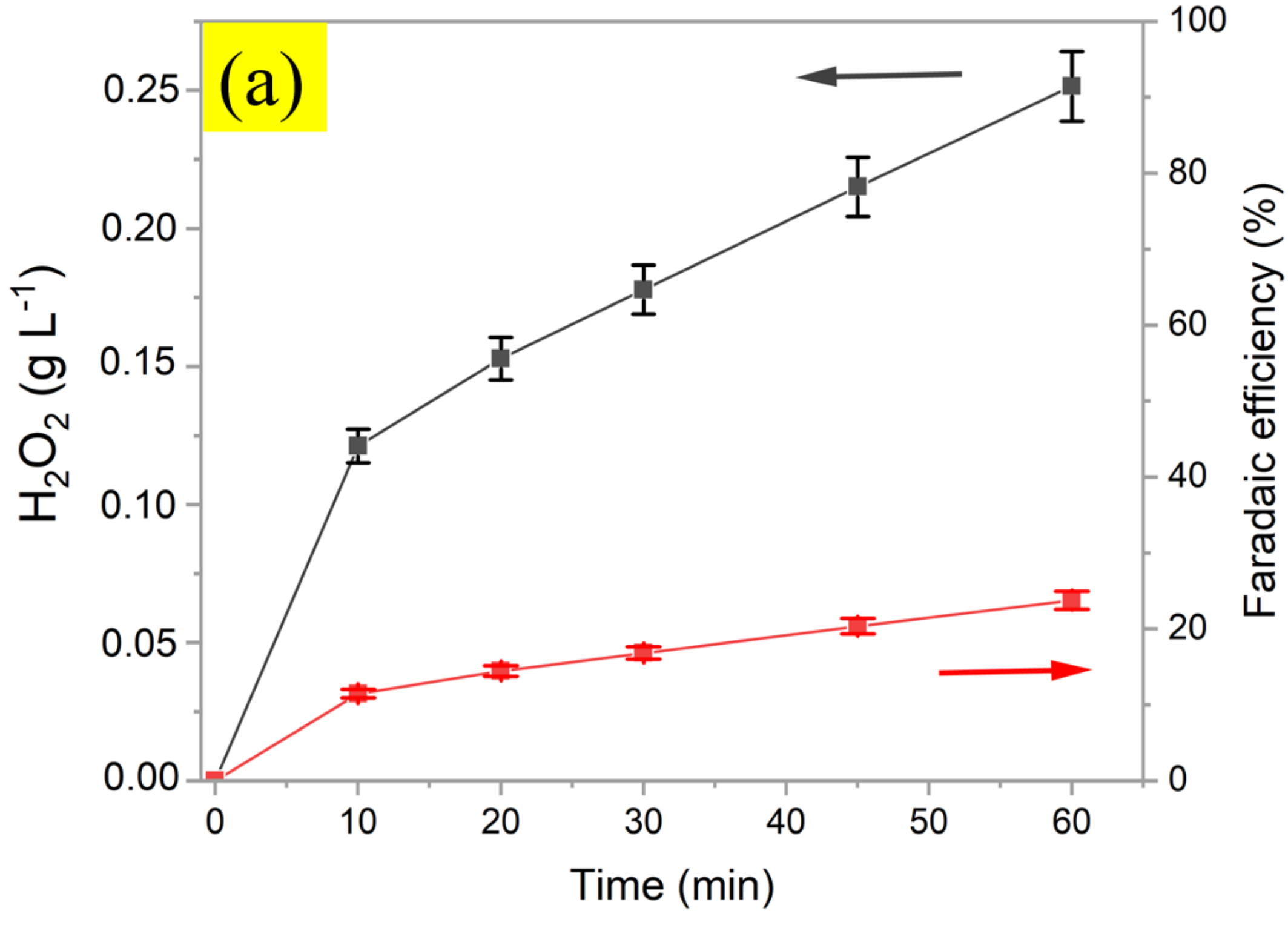
(a)
0.25
0.20
0.15
0.10
0.05
0.00
$H_2O_2$ (g $L^{-1}$)
0
10
20
30
40
50
60
Time (min)
100
80
60
40
20
0
Faradaic efficiency (%)

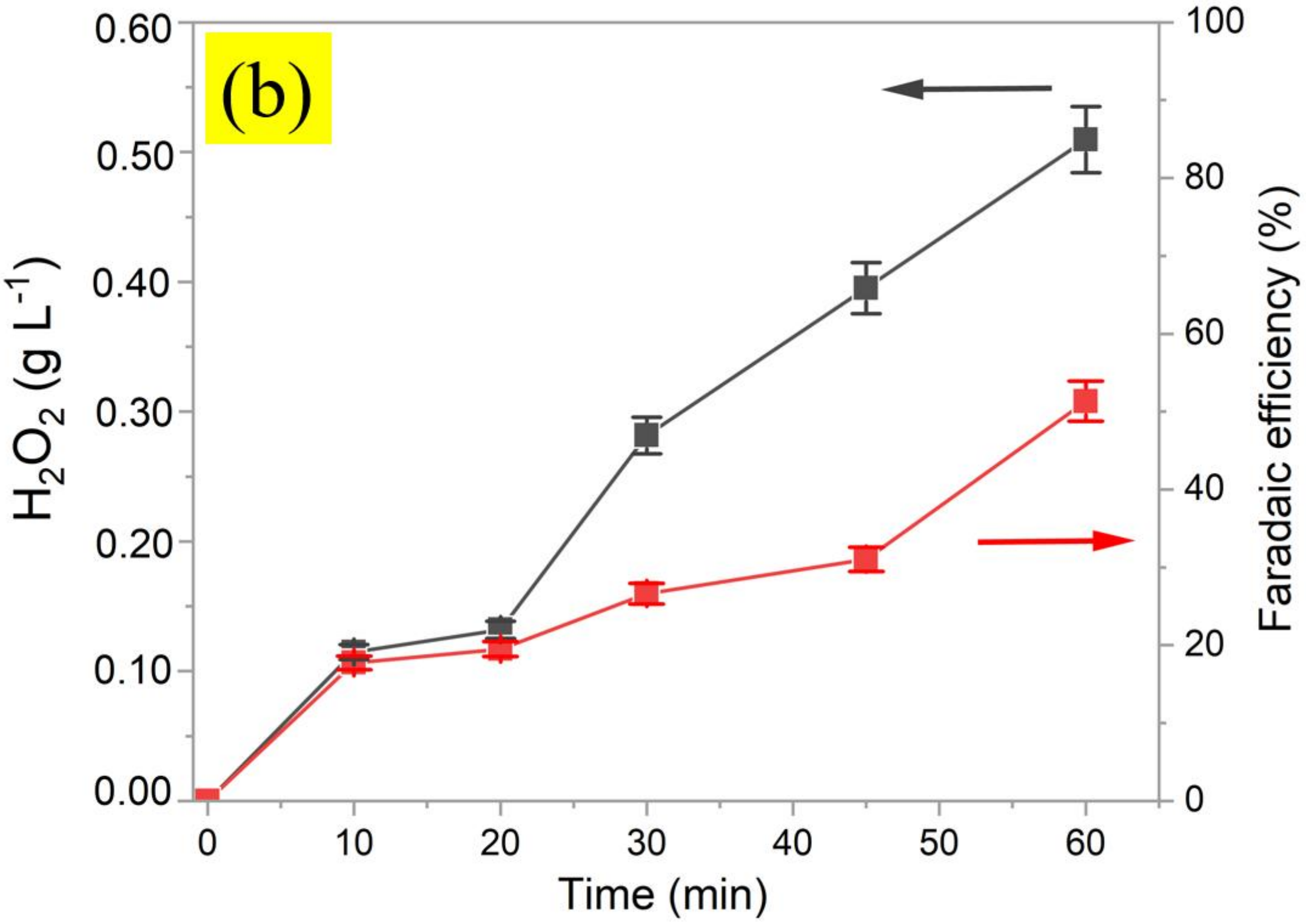
(b)
0.60
0.50
0.40
0.30
0.20
0.10
0.00
$H_2O_2$ (g $L^{-1}$)
0
10
20
30
40
50
60
Time (min)
100
80
60
40
20
0
Faradaic efficiency (%)

**Figure 10**. Evolution of [$H_2O_2$] electrogenerated applied with Vulcan CX-72 (a) and 1% $CeO_2MnOx/C$ (b). Electrolysis time: 60 min. Experimental conditions: $j$ = 137.5 mA $cm^{-2}$; pH = 3.0; [$K_2SO_4$]$_0$ = 0.1 mol $L^{-1}$; anode = platinum.

**Figures 10a** and **10b** show the behavior of electrocatalysts in the electrosynthesis of $H_2O_2$, reaching a maximum concentration of 0.50 g $L^{-1}$ with 1% $CeO_2$–MnOx/C and a double concentration of Vulcan XC-72, under the same experimental conditions. In **Figure 10a**, $H_2O_2$ production increases linearly with time. Still, the FE remains below 20%, indicating that parasitic pathways, including the 4$e^-$ ORR route or surface-mediated decomposition of $H_2O_2$, consume a significant fraction of the applied current. This behavior is characteristic of carbonaceous materials with limited oxygen adsorption strength and insufficient stabilization of the *OOH intermediate, conditions that favor O–O bond cleavage and the water-forming pathway, as widely reported for undoped or weakly functionalized carbons [7,41].

In contrast, the 1% $CeO_2MnOx/C$ (**Figure 10b**) exhibits a substantially higher $H_2O_2$ concentration (>0.50 g $L^{-1}$ at 60 min) and an FE approaching 50%, demonstrating a markedly enhanced selectivity toward the 2$e^-$ ORR pathway. This improvement is consistent with materials featuring optimized surface chemistry, such as oxygen-vacancy-rich metal oxides ($CeO_2MnOx$) or heteroatom-modified carbons, which strengthen $O_2$ adsorption while stabilizing the *OOH intermediate, suppressing the 4$e^-$ route, and minimizing $H_2O_2$ decomposition [42,43]. The steeper production profile and higher FE suggest that the catalyst in **Figure 9b** provides more accessible active sites, better hydrophilicity, and improved electron-transfer properties, all of which enhance $H_2O_2$ accumulation throughout electrolysis. These observations confirm that surface engineering strategies that increase redox activity, oxygen affinity, or defect density are critical for maximizing the efficiency of $H_2O_2$ electrosynthesis under galvanostatic operation.

## 4. CONCLUSIONS

This study aimed to achieve the maximum surface improvements in $H_2O_2$ electrosynthesis with the lowest possible metal loading in carbon-based electrocatalysts. The results showed a significant improvement in the selective electrogeneration of $H_2O_2$ by $CeO_2$ nanowires via the two-electron ORR, providing clear evidence for the initial scientific questions. However, the addition of $CeO_2MnO_x$ reduced this tendency, with the 1% loading being more efficient for ORR than the 3% and 5% loadings.

The applied synthesis routes enabled the preparation of well-defined, small-diameter $CeO_2$ nanowires (8.869 ± 2.269 nm). Microscopic analysis confirmed that Mn was successfully distributed onto the surface of $CeO_2$ nanowires. The investigation of metal versus carbon loading for surface decoration revealed enhanced active-site accessibility, increased surface oxygenated functionalities, and modified wettability, collectively promoting $H_2O_2$ selectivity.

The EPR analysis shows that Mn significantly changes the ceria defect structure. Pure $CeO_2$ has weak signals for oxygen vacancies, but $CeO_2$MnOx displays a 30-fold increase, indicating more vacancies. This increase was related to the MnOx phase, confirming Mn creates a defect-rich lattice. These vacancies improved redox flexibility and likely boost electrocatalytic activity for $H_2O_2$ electrosynthesis.

Among the tested compositions, 1% $CeO_2MnO_x$/C and 3% $CeO_2$/C showed the best catalytic activity, with $H_2O_2$ selectivity of 90% and 70%, respectively. The analysis shows composition-dependent wettability: 1% $CeO_2$ was most hydrophilic (54.13 ± 1.73°), while 5% $CeO_2$ was most hydrophobic (82.54 ± 6.53°), indicating nonlinear effects of $CeO_2$ loading on surface polarity. MnOx addition creates a stable, hydrophilic intermediate regime, similar to Vulcan XC-72, ideal for organic electrochemical applications due to its selectivity for 4-electron reactions.

These findings highlight the potential of non-noble, earth-abundant materials for the sustainable, decentralized production of $H_2O_2$ ($CeO_2$) and for organic oxidation (3–5% $CeO_2MnO_x$), advancing green technologies for environmental remediation and chemical synthesis.

**Acknowledgments**

The authors would like to thank Fundação de Amparo à Pesquisa do Estado de São Paulo (FAPESP, #2021/05364-7, #2021/14394-7, #2022/10484-4, #2022/12895-1, and #2022/15252-4) for the financial support. The authors are also grateful for Coordenação de Aperfeiçoamento de Pessoal de Nível Superior (CAPES) and Conselho Nacional de Desenvolvimento Científico e Tecnológico (CNPq) (#303943/2021-1, #308663/2023–3, #402609/2023–9) for their support.